\documentclass[12pt,preprint]{aastex}

\slugcomment{submitted to ApJ (revised version)}

\shorttitle{}
\shortauthors{Martin, Basri, Pavlenko, Lyubchik}

\begin{document}


\title{Lithium Abundances in Wide Binaries with Solar-Type Twin Components.}


\author{Eduardo L. Mart\'\i n}
\affil{Institute for Astronomy, University of Hawaii at Manoa, 
2680 Woodlawn Drive, Honolulu HI 96822}
\email{ege@ifa.hawaii.edu}

\author{Gibor Basri}
\affil{Astronomy Department, University of California,
    Berkeley, CA 94720}
\email{basri@astro.berkeley.edu}

\author{Yakiv Pavlenko}
\affil{Main Astronomical Observatory of  the Ukrainian Academy of Sciences}
\email{yp@mao.kiev.ua}

\author{Yuri Lyubchik}
\affil{Main Astronomical Observatory of  the Ukrainian Academy of Sciences}
\email{lyu@mao.kiev.ua}



\begin{abstract}
 
We present high-resolution spectroscopic observations of the \ion{Li}{1}
resonance line in a sample of 62 stars that belong to 
31 common-proper motion pairs with twin F or G-type components. 
Photospheric abundances of lithium were derived by 
spectral synthesis analysis. 
For seven of the pairs, we have measured  
large lithium abundance differences.  
Eleven other pairs have components 
with similar lithium abundances.  We cannot determine if the 
remaining 13 pairs have lithium differences  
because we did not detect the \ion{Li}{1} lines, 
and hence we can only provide upper limits to the abundances of both stars. 
Our results demonstrate that  
twin stars do not always share the same lithium abundances. 
Lithium depletion in solar-type stars does not only depend on age, 
mass, and metallicity. This result is consistent with the spread in   
lithium abundances among solar-type stars in the solar-age open cluster M67. 
Our stars are brighter than the M67 members of similar spectral type, making 
them good targets for detailed follow-up studies that could shed light 
on the elusive mechanism responsible for the depletion of lithium during 
the main-sequence evolution of the Sun and solar-type stars. 

\end{abstract}




\section{Introduction}

Lithium has a rich life. It is created in many environments (primordial nucleosynthesis,  
cosmic ray spallation in the interstellar medium, spallation and fusion reactions 
around compact relativistic objects, flares, and red giant thermal pulses), and it is 
destroyed in the interior of stars by collision with protons at a temperature 
of about 2.7$\times 10^6$~K. The Sun has a lithium abundance more 
than two orders of magnitude lower than meteoritic material 
(M\"uller, Peytremann, \& de la Reza 1975), which 
reflects the composition of the presolar nebula. The Sun probably did not 
deplete lithium during the pre main-sequence evolution, but rather during the  
slow main-sequence evolution, although the exact mechanism has not been 
identified yet (Mart\'\i n 1997, 1998). Recent models of lithium depletion in 
solar-type stars suggest that two types of mixing may be at work 
during main-sequence evolution; namely overshooting and rotational 
mixing (Umezu \& Saio 2000). The history of rotational mixing may be strongly 
influenced by the initial conditions of angular momentum distribution in 
the protoplanetary disk. Thus, it is conceivable that there could be 
a connection between the rate of lithium 
depletion by rotational mixing and the presence of companions (stellar or substellar) 
to the stars.

Binary stars with twin components are interesting for understanding binary 
formation and evolution. 
Common proper motion (CPM) pairs of twins allow us 
a test of the validity of stellar evolution models because their separations 
are so large ($>$100 AU) that each member has probably evolved independently. 
If all the properties 
of stars are determined by age, mass, and chemical composition, twins 
should be identical. 
If the lithium abundance of a main-sequence star 
is determined by its age, mass, and chemical composition, pairs of twins 
should show the same lithium abundances. In the early work of Herbig (1965), 
it was already apparent that there was a scatter in lithium abundances 
among pairs of twins. Herbig's study included 53~UMa~A and B, 
where lithium was detected in A but not in B; 16~Cyg A and B, where 
lithium was detected in A but not in B; and 53 Aqr A and B, where lithium 
was detected in both stars.

The system 16~Cyg has received some attention recently. 
The two stars of the pair are nearly identical. The primary has 
V=5.96, T$_{\rm eff}$=5785~K and log~g=4.28. The secondary has 
V=6.20, T$_{\rm eff}$=5747~K and log~g=4.35. Despite those similarities, 
the members of this pair have two fundamental differences. 
King et al. (1997) have shown that the primary has a lithium abundance 
higher than that of the Sun (logN(Li)=1.27$\pm$0.05 in the customary scale 
of log~N(H)=12), while the secondary has a much lower lithium abundance 
(logN(Li)$<$0.60). The large difference in lithium abundance is surprising 
because both stars should be the same age and have very similar masses. 
Another significant difference 
between the two stars is that the secondary harbors a giant planet  with 
a minimum mass of 1.5 Jupiters, an eccentricity of e=0.63 and a period 
of 800.8 days (Cochran et al. 1997). On the other hand, no giant planet has yet 
been detected around 16~Cyg~A. However, such nondetection does not rule out the presence 
of giant planets with long periods or smaller planets with short periods. 

Gonzalez \& Laws (2000) and Ryan (2000) have studied the distribution of 
lithium abundaces among planet-harboring stars, and have reached opposite conclusions. 
While Gonzalez \& Laws claim that lithium is overdepleted in stars with giant planets, 
Ryan claims that it is normal. 
The disagreement comes from the choice of stars to be compared 
with the host stars of extrasolar planets. This is a difficult task because the ages of 
most field stars are not well known. 

Pairs of twins offer the possibility of comparing stars that are coeval, and have nearly 
the same mass and metallicity. Thus, we may expect that the systematic study of binaries 
may shed light on whether there is any connection between the presence 
of giant planets and lithium depletion. Gratton et al. (2001) have recently reported 
an abundance analysis for six wide binaries. They find that the primary star of 
the binary HD~219542 has higher iron and lithium abundances than the secondary star, 
which leads them to suggest that the primary has ingested a planet. 

In this paper we report  
lithium abundances for a sample of 31 visual binaries. The targets have 
been chosen to be as similar as possible to 16~Cyg in spectral type and magnitude. 
The paper is organized as follows: 
In Section 2 we describe the observations. In Section~3 we present the abundance analysis. 
In Section~4 we discuss each pair of twins individually when we could detect the \ion{Li}{1} line 
in at least one of the stars, and we compare them with lithium abundances in open clusters and 
theoretical models. Section~5 contains the discussion, and in Section~6 we present our conclusions.

\section{Observations}

We selected our targets from the list of CPM pairs compiled 
by Halbwachs (1986). We filtered the sample by requiring 
that the difference in V magnitudes should be 
less than 0.4, and the spectral type should be between F8 and K0. 
The declination of the stars had to be north of $-$30 degrees. 
Our sample is listed in Table~1. 

Observations were obtained during several observing runs at Lick and Keck observatories 
between January 1998 and March 2001. The binaries 
BD+08~4386, BD+69~0993, BD+74~0718, and BD+76~0835 were observed at Keck. 
All the other binaries were observed at Lick. The observing log is given in Table~1. 

The Lick observations were carried out using the Hamilton echelle spectrograph on 
the Shane 3-meter telescope. This instrument provides 92 orders covering the  
spectral range from 490~nm to 890~nm at a resolution of 48,000. A standard reduction was performed 
(Valenti 1994), which included background subtraction, cosmic ray removal, and flat 
fielding.  Wavelength calibration was made using ThAr lamp exposures. 

The Keck observations were obtained 
with the HIRES echelle spectrograph (Vogt et al. 1994). 
The resolution was 67,000 and the coverage was from 600 to 850~nm. 
Data reduction was performed using IDL routines similar to those used for the 
Lick data.

\section{Analysis}

We have compiled the available photometry in the literature for our program stars 
and we provide it in Table~2. The $V$ magnitudes, $B-V$ colors, and spectral types were taken 
from SIMBAD, except for the particular cases listed in Section IV, and for the notes to 
Table~2. 
This photometric database is rather heterogeneous. Some of our stars 
are included in the 2MASS second incremental release point source catalog (PSC). 
Their 2MASS $J-K$ colors are included in Table~2.    
Temperatures were derived using the empirical scale of Alonso, Arribas 
\& Mart\'\i nez-Roger (1996) for solar metallicity. 
The temperatures obtained from the $J-K$ colors are given in Table 3. 
Typical error bars in $J-K$ colors are $\pm$0.03, corresponding to $\pm$180~K in T$_{\rm eff}$. 
These temperatures were derived to compare with our, presumably more reliable, 
temperatures obtained from synthetic spectra fitting to the observed spectra 
in the lithium region. This comparison is shown in Figure~1 for 26 stars that have available 2MASS 
$JK$ photometry. The mean difference between the T$_{\rm eff}$ derived from $J-K$ and from fits is 
$-$65~K and the standard deviation is 270~K. We conclude that there is no significant systematic 
deviation of the T$_{\rm eff}$ derived from $J-K$ colors with respect to those derived from 
synthetic fits.  

For each spectrum we determined the continuum level 
using the DECH20 software (Galasutdinov 1992). To detect the 
real continuum level, we smoothed the observed spectra  
 using three loops of a 3-point smoothing boxcar. 
Then the continuum points were fitted by splines to use in the fine analysis procedure. 
The S/N ratios per pixel given in Table~1 were obtained by measuring the standard deviation of 
the normalized continuum in the spectral range 669.3-669.5~nm. 
We carried out the computations using the LTE spectral synthesis
program WITA6 (Pavlenko 2000), which is a modified version of 
the program used by Pavlenko et al. (1995).
Our computations of synthetic spectra were carried 
out using plane-parallel model atmospheres in LTE, 
with no energy divergence. The model atmospheres were taken from 
Kurucz's grid available on the web. 

We used metallicities in the range 
[m/H]=0.0 to [m/H]=-0.2 for fitting the observed 
spectra\footnote{We refer to the solar abundances given by 
Grevesse (1984) using the convention [m/H]=0} except for 
HD 54046 and HD 54100, which are known metal poor stars (Norris 1986).  
Chemical equilibrium was computed for  the  mix  of  $\approx$100 
molecular species, with data taken from Tsuji (1973). 
ATLAS9 (Kurucz 1993) grid of continuum opacity sources were used in our 
computations. 
We used VALD (Piskunov et al. 1995) for the atomic line list and 
damping constants. For some lines the damping data are 
missing; in that case we used Uns\"old (1955) approach to compute 
them. We had to increase the log~gf of the FeI line at 670.510~nm 
by a factor of 2 (from 0.0319 given by VALD to 0.0640)  to 
find good fits to the observed spectra of 
the Sun as a star (Kurucz et al. 1984). Computations were carried out for a fixed 
microturbulent velocity of $V_t$=2~km~s$^{-1}$  
and a Voigt profile for every absorption line.    

Synthetic spectra were computed with a step of 0.002~nm. Instrumental 
profile and macroturbulence broadening effects were 
modeled by convolution with a Gaussian of half-width in the range 0.010~nm 
to 0.018~nm. 
For most stars we could desribe the profile with instrumental 
broadening only, without any rotation. That provides 
upper limit of vsin{\it i}=3~km~s$^{-1}$. No enhanced van-der Waals broadening 
was used. 
For two stars rotational broadening was taken into account 
(HD~6872~A and BD+65~1044; vsin{\it i}=14~km~s$^{-1}$).  
For each star we found the best-fitting synthetic spectrum using a least-squares 
minimization algorithm. Figures 2, 3 and 4 show typical examples of synthetic fits 
to the echelle spectra. 

For each LTE lithium abundance, we determined NLTE abundance 
corrections usind the grid of curves of growth of Pavlenko \& 
Magazz\`u (1996).  In Table~3 we summarize our results. 
We give the temperatures, metallicities and lithium abundances (including 
NLTE corrections) that we used in the best fits to the observed spectra. 
Equivalent widths of the \ion{Li}{1} line measured by direct integration of the line 
profile in the observed spectrum are given in Table~3 as well. They can be used 
to derive lithium abundances from curve of growth (COG) computations. 
For the stars with 2MASS photometry we have obtained COG Li abundances using the 
temperatures derived from the $(J-K)-T_{\rm eff}$ calibration of Alonso et al. (1996) 
and the WITA code. 
Those abundances are listed in Table~3 for comparison with the abundances obtained 
from the synthetic 
spectra. Figure~5 illustrates this comparison for the same 26 stars that were used 
in Figure~1. The mean difference between the lithium abundances derived using these 
two methods is 
$logN(Li)_{\rm COG} - logN(Li)_{\rm fit}=-0.08$ dex, and the standard deviation is 0.17~dex. 
We conclude that there is no significant systematic discrepancy between the two methods of 
obtaining 
lithium abundances. We adopt the uncertainty of 1~$\sigma$=0.17~dex in our absolute lithium 
abundances. In our computations we have assumed that our stars are dwarfs. 
A variation of log~g $\pm$0.5 dex provides a maximum $\Delta$logN(Li)$<$0.1 dex.

\section{Comments on Individual Stars}

The main goal of this paper is to determine whether stars with the same age and initial 
chemical composition, and nearly equal masses, have depleted lithium at the same rate. 
We limit our discussion to pair of twins for which we could detect the \ion{Li}{1} line in  
at least one of the stars. When we could not detect the \ion{Li}{1} line in any of the stars, 
we could not tell if the lithium abundances are different or similar, and therefore those 
pairs are not discussed any further.  After a literature search using SIMBAD, 
we found that some of the pairs deserve individual comments: 

\begin{itemize} 
\item{HD~6872 A and B:  
Oblak \& Chareton (1980) and 
Duncan (1984) included HD~6872~A and B among a sample of stars that are sufficiently evolved 
for the ages to be estimated from T$_{\rm eff}$ and absolute magnitudes using the displacement 
from the zero-age main-sequence (ZAMS). Both groups obtained uvby$\beta$ photometry and derived 
T$_{\rm eff}$ of 6380~K (Oblak \& Chareton) and 6300~K (Duncan) for the primary. 
For the secondary, they gave 6251~K (Oblak \& Chareton) and 6084~K (Duncan). 
The discrepancy in T$_{\rm eff}$ is mainly due to their use of different equations of 
$b-y$ index versus temperature. 
Soderblom, Duncan \& Johnson (1991) converted the $b-y$ color of Duncan to $B-V$ color. 
In Table~1, we adopt the Soderblom et al. color, which is different from the color provided by SIMBAD. 
Our best fit temperatures are 6250~K for both stars. 
We find good synthetic fits to our spectra with the same T$_{\rm eff}$ for both stars 
(Figure~3). 
The fainter star has a higher lithium abundance than the hotter star by 
at least 0.8 dex. Adopting the T$_{\rm eff}$ of either 
Oblak \& Chareton or Duncan does not change the conclusion that the lithium abundances of these 
two stars are significantly different.  
The primary star has broader lines  than the secondary. Our best fit to the spectrum of 
HD~6872 A has rotational broadening of vsin{\it i}=14 km~s$^{-1}$.} 

\item{HD 39274 and HD 39275: The lithium abundance difference between these two stars is 0.1 dex, 
which is less than the 1~$\sigma$ uncertainty in our analysis. 
HD~39274 is a binary with separation 0.225 arcsec 
(McAlister et al. 1989). }  

\item{HD 54046 and HD 54100: Metal poor stars. 
Norris (1986) reported [Fe/H]=-0.6 and Eggen (1998) gave [Fe/H]=-0.45 based on photometric colors.}

\item{HD~8624 and HD~8610:  
Tokovinin (1999) found that HD~8624 
is a double-lined spectroscopic binary with a period of 14.9 days. HD~8610 
has a K-type companion, and therefore the system is a quadruple. 
Tokovinin gave V=8.83 for both components of the HD~8624 close system, and inferred 
a spectral type of G3. }

\item{HD 98744 and HD 98745:  
HD 98745 has a lithium abundance 1 dex or more higher than HD 98744. 
Duncan (1984) found T$_{\rm eff}$=6050~K and 6039~K for HD 98744 and HD 98745, respectively. 
Our best fit T$_{\rm eff}$ is 6250~K for both stars. 
Adopting the T$_{\rm eff}$ of Duncan does not change the conclusion that the lithium abundances of these 
two stars are significantly different.}

\item{HD~167215 and HD~167216: 
In Table~2 we give the $B-V$ colors given by SIMBAD, because they are 
more consistent with the temperatures that we find in our analysis, than 
the $B-V$ colors of Soderblom et al. (1991). These latter authors derive 
T$_{\rm eff}$ of 5984~K and 6040~K for 
HD~167215 and HD~167216, respectively. We find that we need hotter temperatures  
to fit the spectra in the lithium region. We also need subsolar metallicities. 
The lithium abundance difference is 0.1 dex. The \ion{Li}{1} resonance line of HD~167215 is clearly stronger 
than that of HD~167216. According to Soderblom et al., HD~167215 is more evolved 
above the ZAMS than HD~167216. The gravity of HD~167215 may be somewhat lower than 
that of HD~167216, and this could explain the difference in \ion{Li}{1} line strengths.}  
  
\item{HD~224984 and HD~224994: 
These two stars have nearly identical spectra.  HD~224994 itself is a close binary 
(separation=0.24 arcsec in 1988; McAlister, Hartkopf, \& Franz 1990), so the system 
is at least triple. The two stars have very similar spectra, including the \ion{Li}{1} lines. 
We obtain lithium abundances of logN(Li)=2.3, which is about 0.8 dex below the cosmic value. 
Therefore, 
this is an example of a pair of twins that have depleted the same amount of lithium. 
The presence of a low-mass star companion at about 20~AU of HD~224994 has not changed 
the lithium depletion rate of this star with respect to HD~224984. }
\end{itemize}
 
\section{Discussion} 

Lithium abundance is very sensitive to the mass (and thus T$_{\rm eff}$) of stars. 
In the temperature range of our program stars (6500-5250~K), lithium abundances generally 
decrease with decreasing T$_{\rm eff}$, although at the hot end of this temperature range, 
the F-type lithium dip starts to kick in (Boesgaard \& Tripicco 1987). 
If the lithium abundances 
of the components of pairs of twins were dominated by the mass dependance of lithium depletion, 
we would expect a correlation between the difference in T$_{\rm eff}$ and the difference in 
logN(Li), in the sense that the cooler components should show systematically lower lithium 
abundances than the hotter ones. Figure~6 shows that such correlation is not present in the data. 
We conclude that the observed pattern of lithium abundances in pairs of twins cannot be explained 
solely with the dependance of lithium depletion on stellar mass. 
 
We have found five pairs with nearly identical T$_{\rm eff}$ and nearly identical logN(Li), 
and four pairs with nearly identical T$_{\rm eff}$ and very different logN(Li). 
Hereafter we call these latter four pairs "16~Cyg~analogs", namely:  
HD~6872 A/B, BD+60~0269/BD+60~0271, BD+13~2311~A/B, and 
HD 98744/HD 98745. The common characteristics of "16~Cyg~analogs" is that they are   
pairs of twins with lithium abundance difference between the two stars of the pair 
that exceeds 0.5 dex  
(which corresponds to 3~$\sigma$ significance in our synthetic fit analysis) 
and difference in T$_{\rm eff}<$200~K. 

Figure~7 illustrates the comparison between the pairs of twins and the Hyades cluster in a 
T$_{\rm eff}$ versus logN(Li) diagram. The Hyades single star members define a tight relationship 
between T$_{\rm eff}$ and logN(Li). In general the twins do not follow the Hyades relation. 
A possible explanation for this discrepancy could be that 
there are many tidally locked binaries (TLB) hidden among the twins.  
We do not know of any example of a TLB in our sample. 

HD~8610 has a lithium abundance similar to Hyades members with the same T$_{\rm eff}$. 
The lithium-T$_{\rm eff}$ locus in the Hyades declines steeply from 5500~K to 5250~K, so one would expect 
that HD~8624 should have a lithium abundance much lower than that of HD~8610, which 
is contrary to what we have found. HD~8624 
is a double-lined spectroscopic binary with a period of 14.91 days (Tokovinin 1999) and 
an eccentricity e=0132. Even though it is not strictly a TLB because the orbit is not completely circularized, 
it may have preserved lithium in a manner 
similar to the TLBs studied by Barrado y Navascu\'es et al. (1997), which show a trend of higher 
lithium abundances when compared to single stars. However, TLBs are relatively rare among the 
general population of solar-type stars. A program to monitor the radial velocity of 10 of our 
pairs has been started using HIRES with the iodine cell. There are already enough data for 
16 of the stars to find TLBs, but none has been seen (Marcy 2002, private communication). 
We conclude that it is very unlikely that TLBs can account for most  
of the discrepancy between the pairs of twins and the Hyades. 

The average age of the pairs of twins is probably larger than the Hyades cluster age ($\sim$600~Myr). 
It is informative to compare the twins with the well-studied  
solar-age cluster M67. In Figure~7 we compare the pairs of twins with the 
members of M67.  The M67 stars show a large spread of lithium abundances at a given 
T$_{\rm eff}$, which is similar to the behavior exhibited by our binaries. The evolution of 
lithium from the Hyades to M67 implies a mechanism of lithium depletion during the main-sequence 
lifetime of solar-type stars that is not very mass dependent. We have not found in the literature 
any models that can explain the evolutionary pattern of lithium abundances implied by the 
comparison between the Hyades and the M67 solar-type stars. We conclude that there is currently 
not a good theoretical explanation for the lithium abundances observed in our pairs of twins 
and the open clusters. In future papers we plan to tackle this problem observationally via 
searches for connections between the lithium abundances in the pairs of twins and 
the presence of companions (stellar or substellar), chromospheric activity, rotation, and 
the abundances of other chemical elements.  
At the moment it would be premature to suppose that the presence of 
planets had anything to do with the spread in lithium abundances.

\section{Final Remarks}

We report high-resolution spectroscopy observations of 62 solar-type stars in 
31 common proper motion pairs of twins. These binaries are sufficiently separated 
that the evolution of each star is not expected to be influenced by its  
stellar companion. Lithium abundances have been derived from 
theoretical fits to the observed spectra using model atmospheres. 
A comparison with COG abundances derived from \ion{Li}{1} equivalent widths and temperatures 
obtained from near-infrared colors shows that the absolute  
lithium abundances are reliable to within 0.17 dex (1~$\sigma$).  

We have found seven pairs of twins with large lithium abundance differences. 
The pattern of lithium abundances in our sample of twins seems to be midway between those 
of the Hyades (tight lithium-T$_{\rm eff}$ relation) and M67 (large lithium spread for 
any T$_{\rm eff}$) open clusters. 
We have not found any evolutionary models in the literature 
that provide a satisfactory explanation for the 
observed spread in lithium abundances in solar-type stars. We plan to follow up with additional 
studies of our sample of twins to obtain an overall understanding of all the effects 
that may come into play in the evolution of solar-type stars.

\acknowledgments

This paper is based on data collected with the 3-m Shane telescope at Lick Observatory, California, 
and the Keck~I telescope at the W. M. Keck Observatory, Mauna Kea, Hawaii. The Keck observatory is 
operated as a scientific partnership among the California Institute of Technology, 
the University of California, and the National Aeronautics and Space Administration. 
The Observatory was made possible by the generous financial support of the W. M. Keck 
Foundation.  
This research has made use of the SIMBAD database, operated at CDS, Strasbourg, France. 
The authors wish to extend special thanks to those of Hawaiian ancestry on whose sacred 
mountain of Mauna Kea we are privileged to be guests. Without their generous hospitality, 
the Keck telescope observations presented therein would not have been possible.  
We thank Louise Good for correcting the English of the manuscript, and Geoff Marcy 
for communicating results of the radial velocity program for a subset of our sample. 
We are also indebted to Ann Boesgaard and David Barrado 
y Navascu\'es for sending data files for lithium abundances of M67 stars, 
and Nina Kharchenko for astrometric data.  
The project was partly supported by a SRG grant of AAS to YP.

\clearpage


\begin{figure}
\plotone{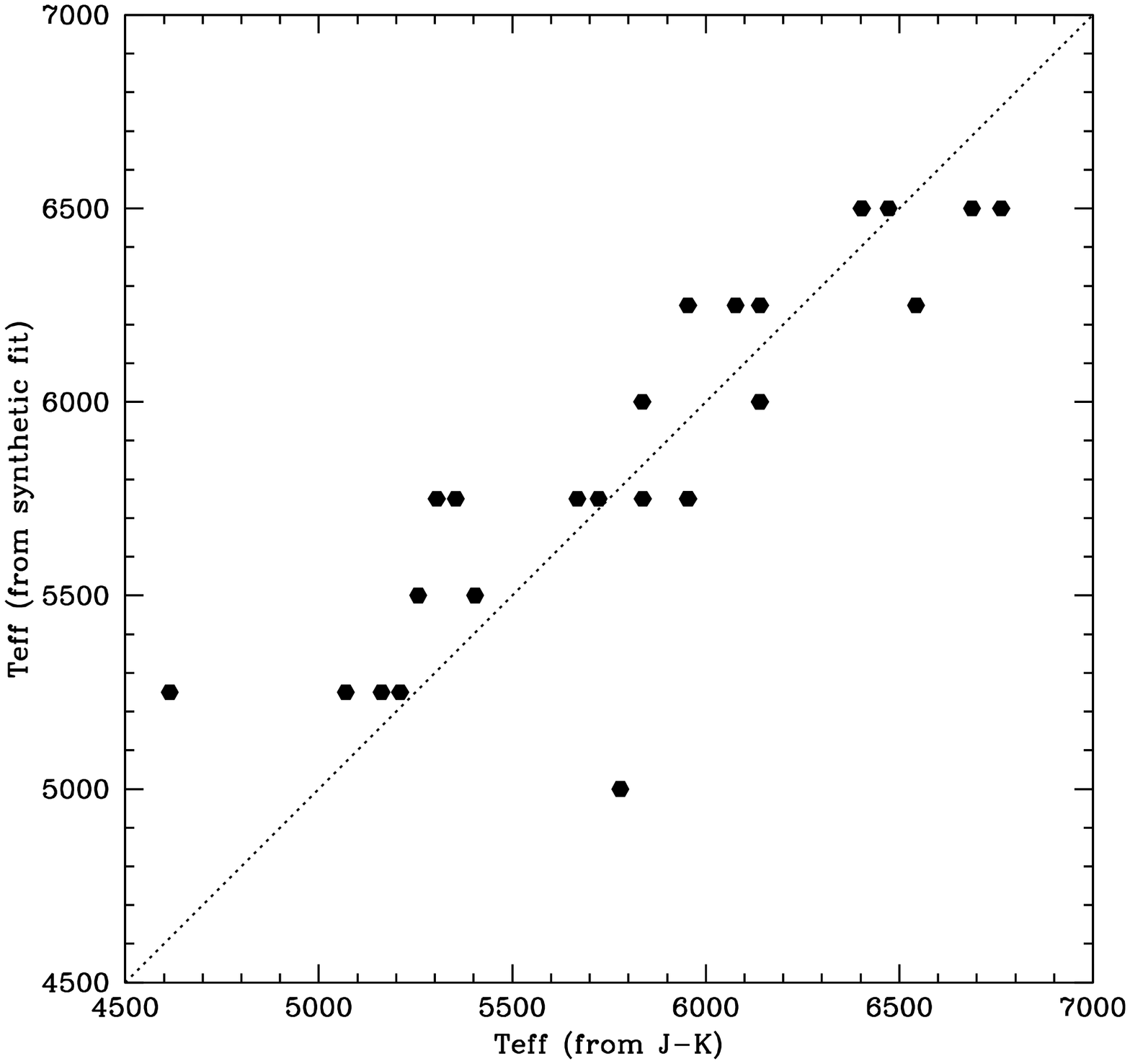}
\caption{Comparison of effective temperatures obtained from $J-K$ colors and from the best 
synthetic fits to the observed spectra.}
\end{figure}

\begin{figure}
\plotone{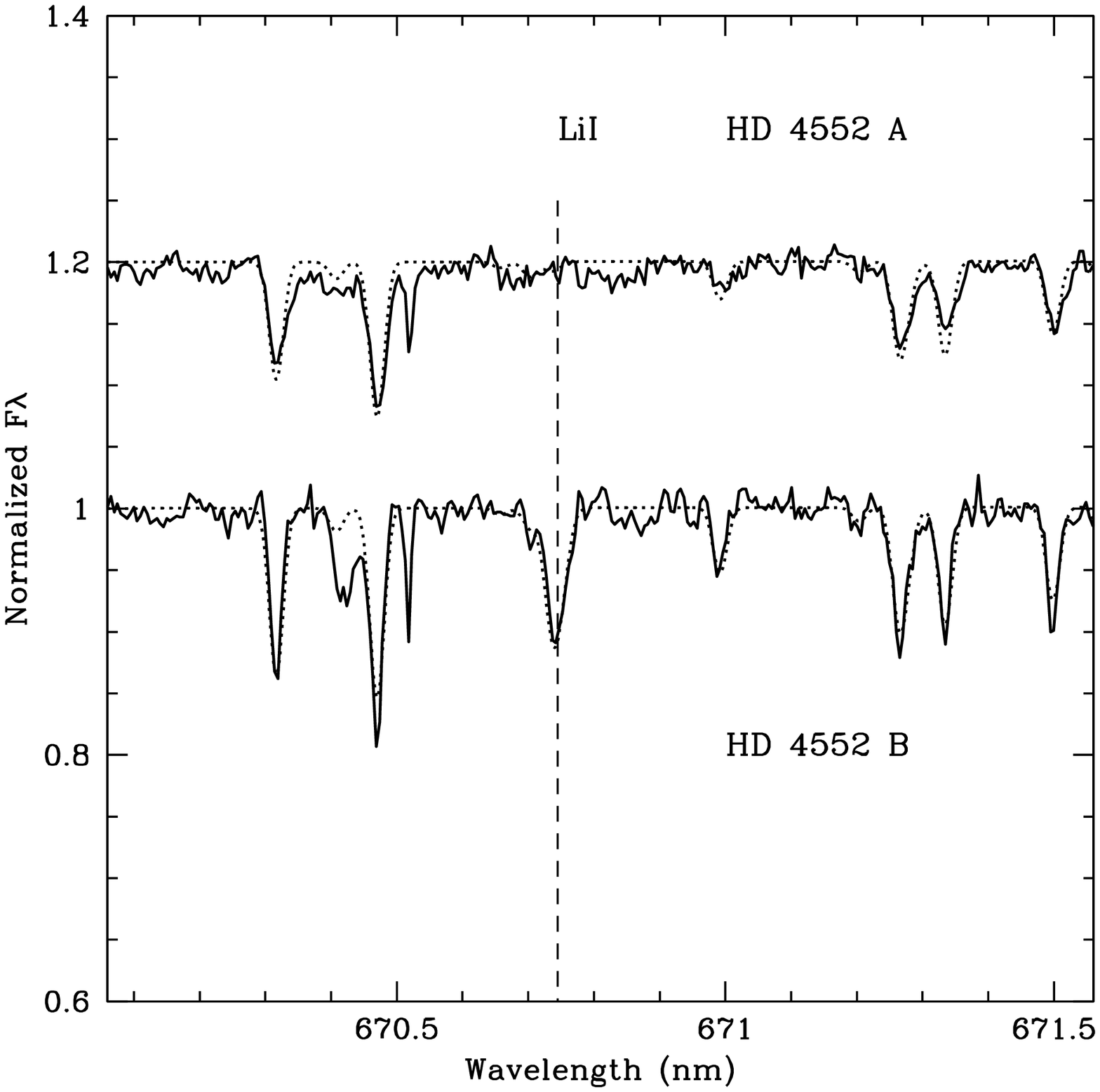}
\caption{Solid lines: Lick Hamilton echelle spectra of the CPM pair HD 4552~A and B. 
Dashed lines: Synthetic spectra for T$_{\rm eff}$=6000~K, [m/H]=$-$0.2, and 
logN(Li)=1.6 (above), and for T$_{\rm eff}$=5750~K, [m/H]=$-$0.2, and 
logN(Li)=2.3 (below).}
\end{figure}

\begin{figure}
\plotone{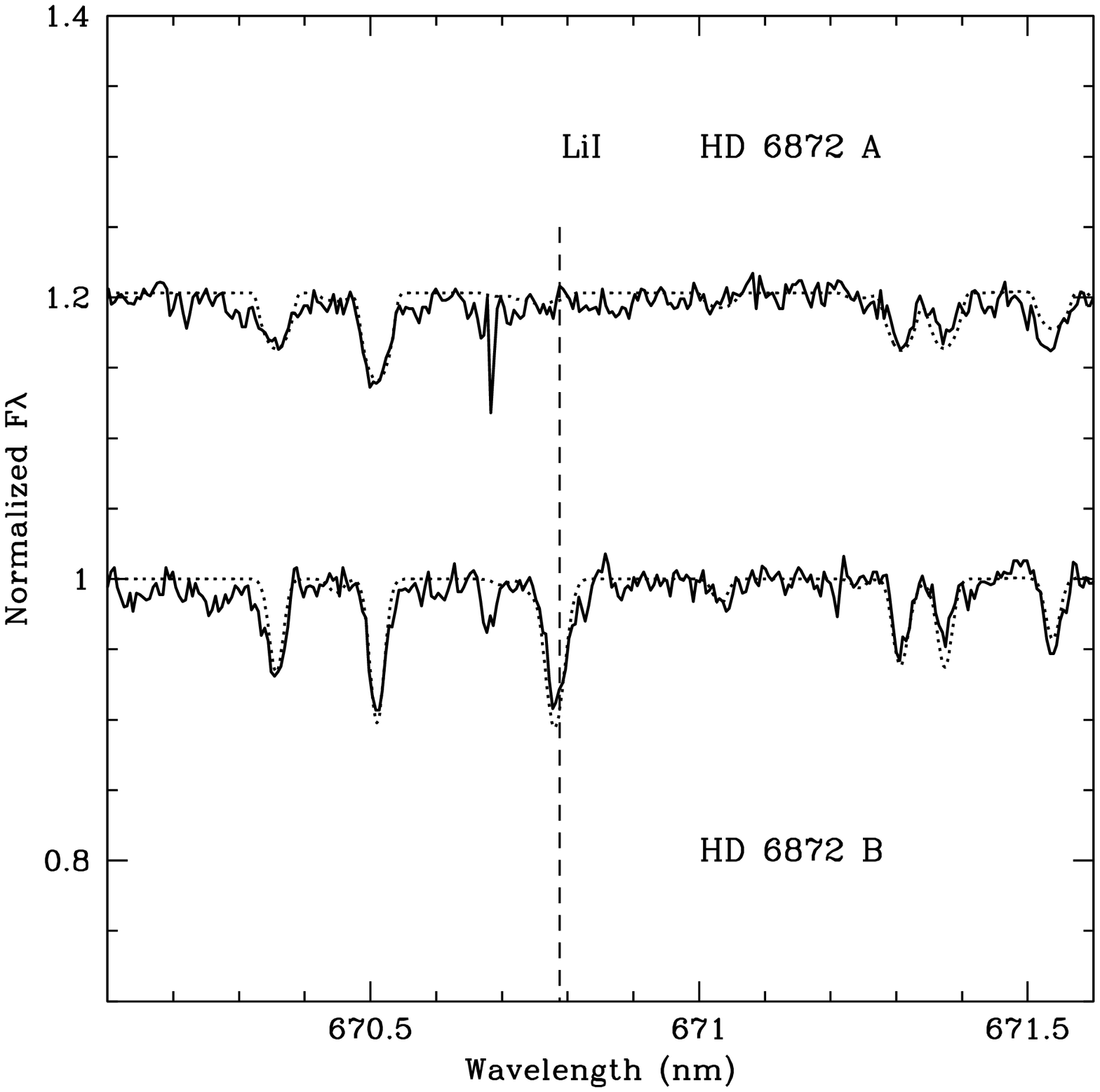}
\caption{Solid lines: Lick Hamilton echelle spectra of binary HD~6872 A and B. 
Other name for HD~6872~B is BD+12~0090. 
Dashed lines: Synthetic spectra for T$_{\rm eff}$=6250~K, [m/H]=$-$0.2, and 
logN(Li)=1.8 (above), and for T$_{\rm eff}$=6250~K, [m/H]=$-$0.2, and 
logN(Li)=2.6 (below).}
\end{figure}

\begin{figure}
\plotone{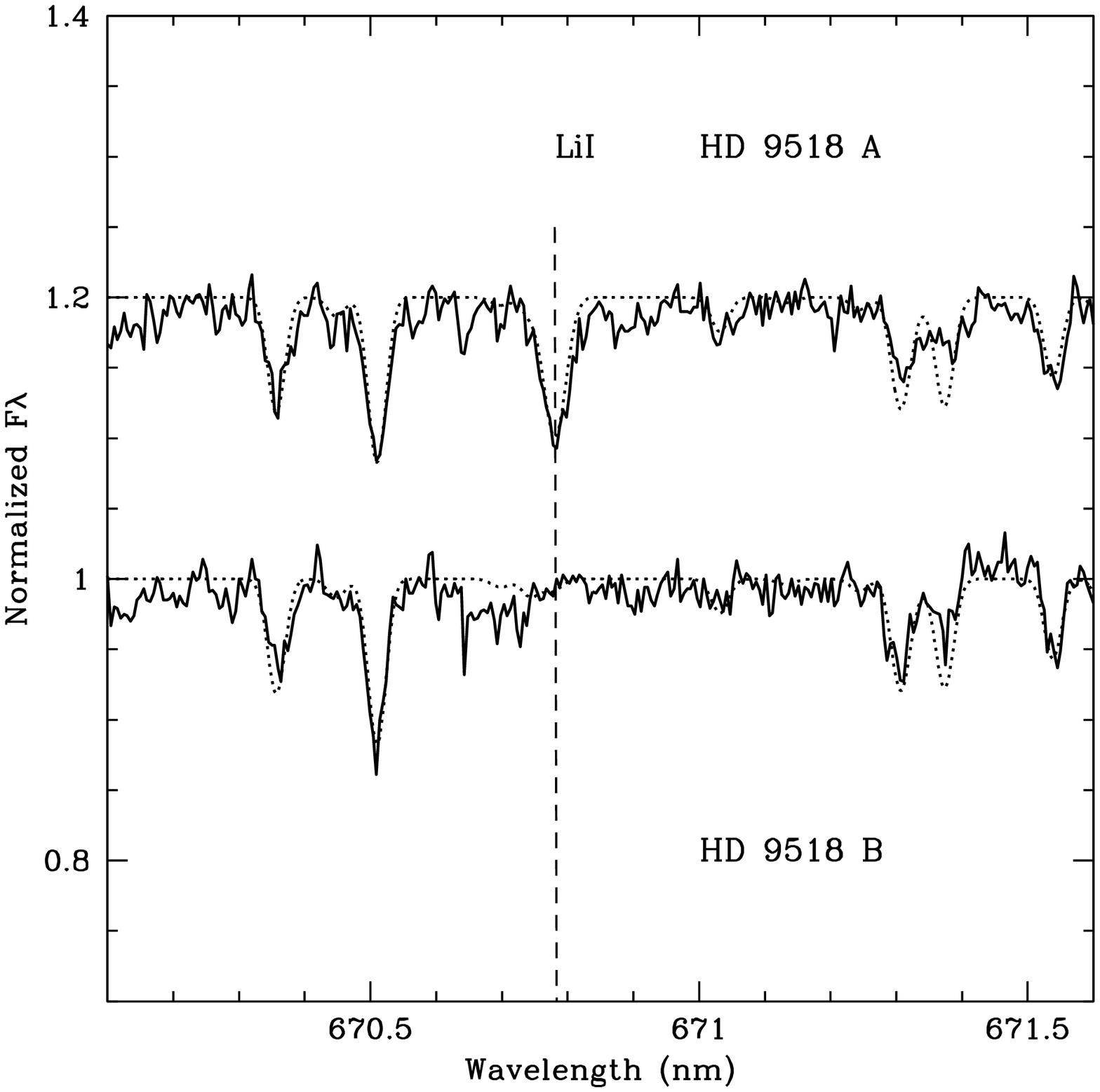}
\caption{Solid lines: Lick Hamilton echelle spectra of the CPM pair 
HD~9518~A (BD+60~0269) and B (BD+60~0271). 
Dashed lines: Synthetic spectra for T$_{\rm eff}$=6250~K, [m/H]=0.0, and 
logN(Li)=2.55 (above), and for T$_{\rm eff}$=6250~K, [m/H]=0.0, and 
logN(Li)=1.8 (below).}
\end{figure}

\begin{figure}
\plotone{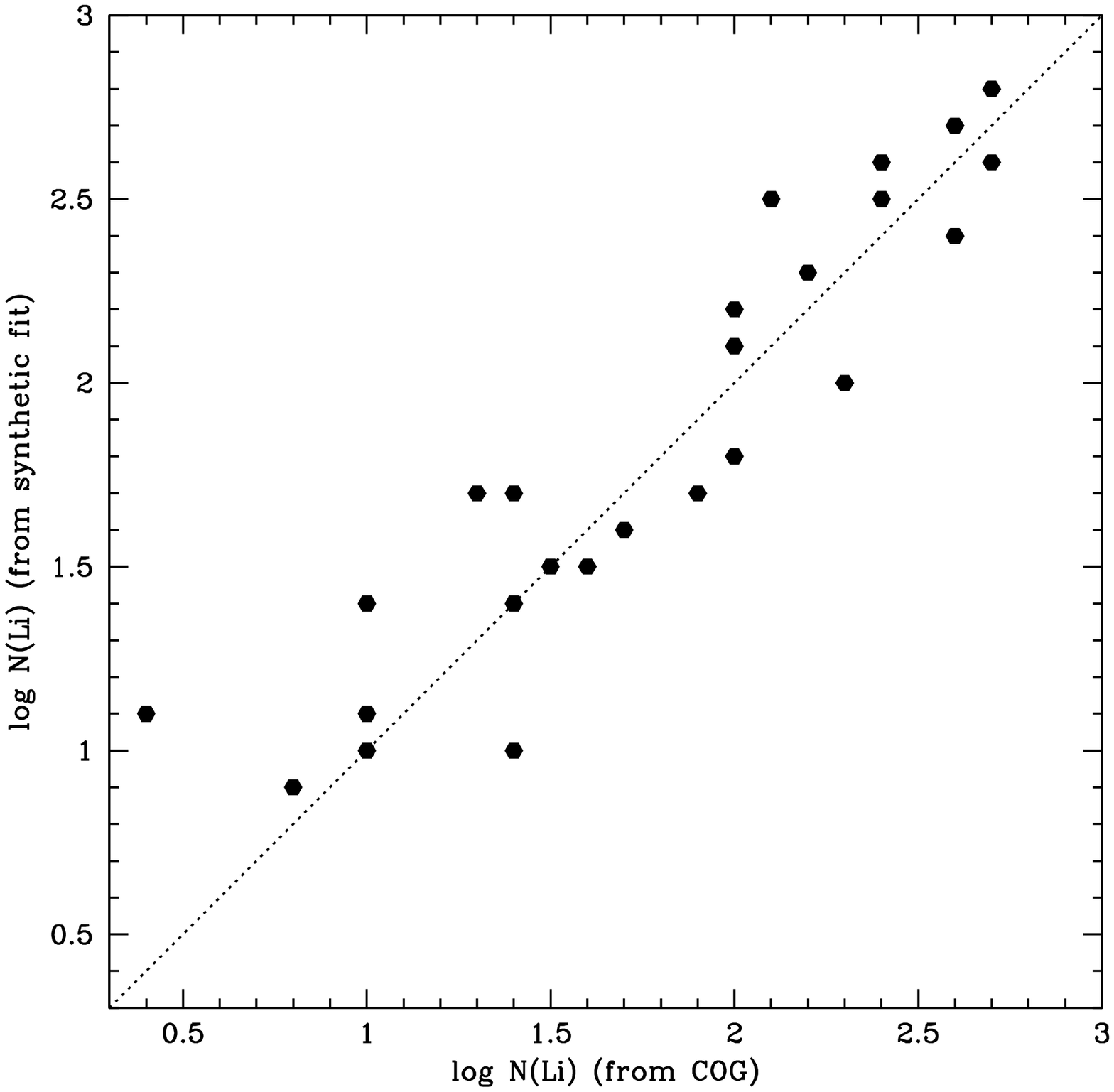}
\caption{Comparison of lithium abundances obtained with $J-K$ colors and COG calculations 
and with the best synthetic fits to the observed spectra.}
\end{figure}

\begin{figure}
\plotone{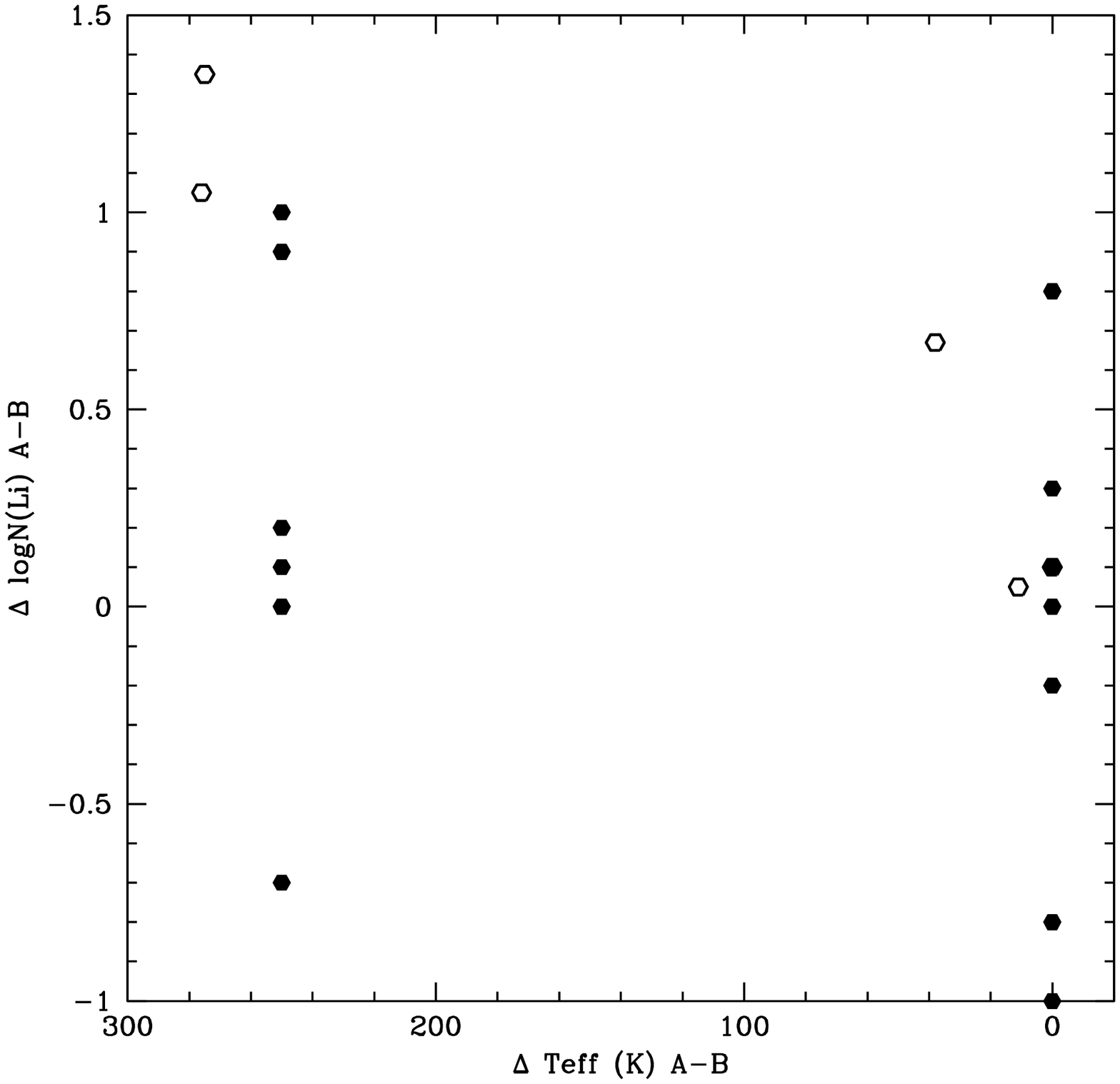}
\caption{Differences in the T$_{\rm eff}$ adopted for pairs of twins versus the differences 
in the lithium abundances derived by us (solid symbols) and by previous investigations (open symbols).}
\end{figure}

\begin{figure}
\plotone{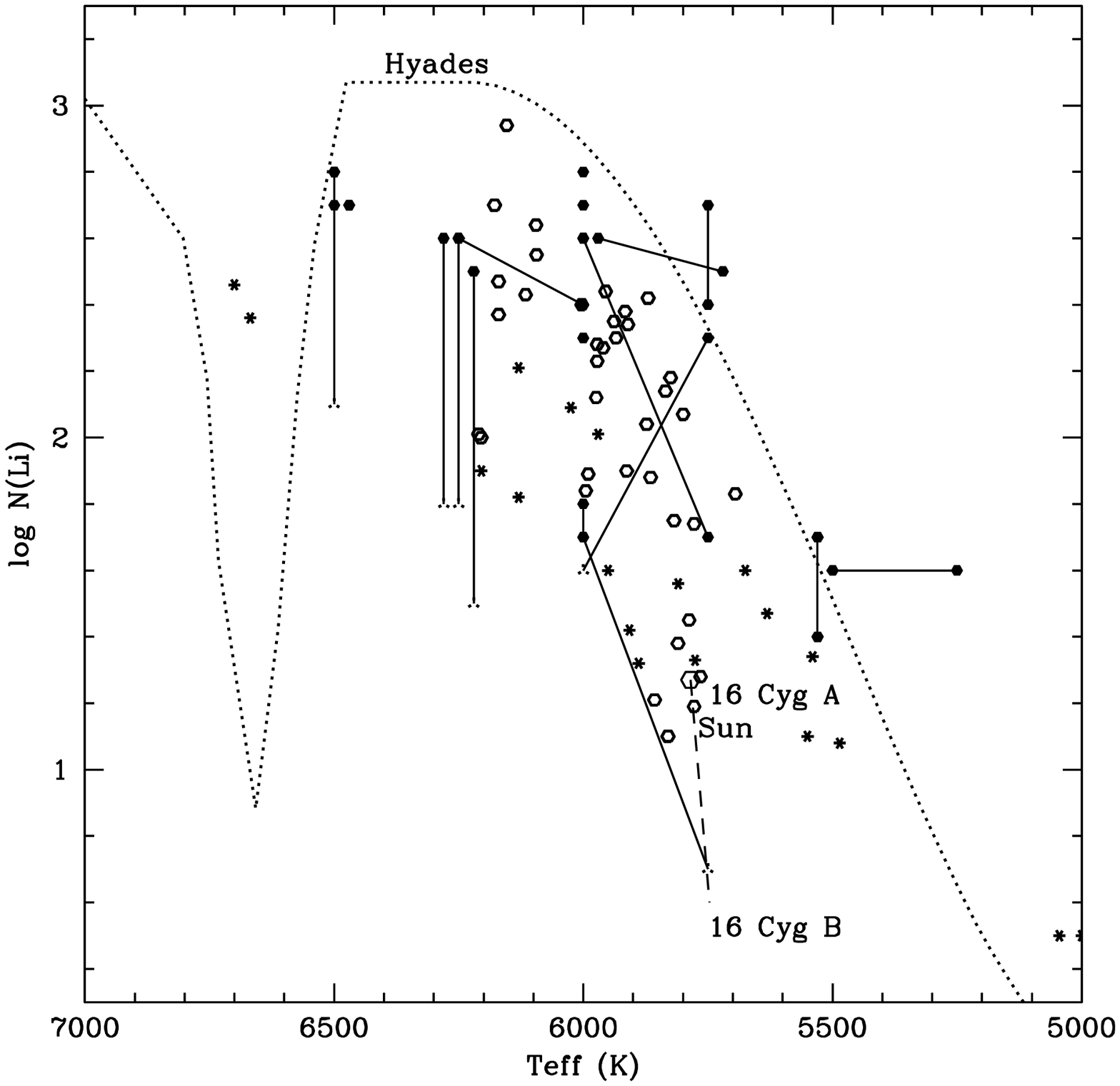}
\caption{Lithium abundances versus T$_{\rm eff}$ for pairs of twins. Our lithium measurements are given 
as solid dots, with the components of each binary linked by a solid line. The effective temperatures of 
some pairs are shifted by 30~K to avoid confusion. 
The components of 16~Cyg 
are denoted with an open dot and a solid line. The locus of lithium abundances in single stars 
members of the Hyades cluster is shown with a dotted line. The Hyades data have been taken 
from Thorburn et al. (1993). 
Lithium abundances for M67 members are denoted with open hexagons for 
detected \ion{Li}{1} lines, and asterisks for upper limits. Data for M67 stars come 
from Hobbs \& Pilachowski 
(1986), Spite et al. (1987), Garc\'\i a L\'opez, Rebolo, \& Beckman (1988), 
Deliyannis et al. (1994), Pasquini, Randich, \& Pallavicini (1997), Jones, Fischer, \& Soderblom 
(1999), Barrett et al. (2001), and Barrado y Navascu\'es, \& Balachandran (2002).}
\end{figure}

\clearpage

\begin{deluxetable}{llllr}
\tabletypesize{\scriptsize}
\tablecaption{Observing Log. \label{tbl-1}}
\tablewidth{0pt}
\tablehead{
\colhead{HD/BD} & 
\colhead{Telescope} & 
\colhead{Date} &
\colhead{Texp (s)} & 
\colhead{S/N} 
}
\startdata
HD~4552      & 3-m Lick    & 1998 Nov 23   &  1500 & 140   \\
BD+12~0090   & 3-m Lick    & 1998 Nov 23   &  1800 & 105   \\
HD~6872~A    & 3-m Lick    & 1998 Sept 4  &  1000  & 140   \\ 
HD~6872~B    & 3-m Lick    & 1998 Sept 4  &  1100  & 130   \\
HD~8624      & 3-m Lick    & 1998 Sept 3  &   600  & 105   \\ 
HD~8610      & 3-m Lick    & 1998 Sept 3  &   600  & 130   \\
BD+60  269   & 3-m Lick    & 1998 Sept 3  &   800  & 110   \\
BD+60~0271   & 3-m Lick    & 1998 Sept 3  &  1400  & 100   \\
HD 31208     & 3-m Lick    & 2001 March 8  &  1800 & 150   \\
BD+07 754    & 3-m Lick    & 2001 March 8  &  1800 & 190   \\
HD 39274     & 3-m Lick    & 2001 March 8  &  2500 & 190   \\
HD 39275     & 3-m Lick    & 2001 March 8  &  2200 & 175   \\
HD 54046     & 3-m Lick    & 1998 Jan 22   &  1800 & 280   \\
HD 54100     & 3-m Lick    & 1998 Jan 22   &  1800 & 340   \\
BD+28~1698   & 3-m Lick    & 1998 Jan 21   &  1200 & 210   \\
BD+28~1697   & 3-m Lick    & 1998 Jan 21   &  1200 & 200   \\
BD+15~2080   & 3-m Lick    & 1998 Jan 22   &  2400 & 210   \\
BD+15~2079   & 3-m Lick    & 1998 Jan 22   &  1800 & 250   \\
BD+34 2091 A & 3-m Lick    & 2001 March 9  &  3000 &  90   \\
BD+34 2091 B & 3-m Lick    & 2001 March 9  &  3600 &  95   \\ 
HD 92222 A   & 3-m Lick    & 2001 March 8  &  3600 & 140   \\
HD 92222 B   & 3-m Lick    & 2001 March 8  &  3600 & 135   \\
BD+13 2311 A & 3-m Lick    & 2001 March 9  &  5130 & 225   \\
BD+13 2311 B & 3-m Lick    & 2001 Jan 22   &  4400 & 220   \\
HD 98744     & 3-m Lick    & 2001 March 8  &  4000 & 175   \\
HD 98745     & 3-m Lick    & 2001 March 8  &  4500 & 175   \\
HD~111484~A  & 3-m Lick    & 1998 Jan 21   &  1800 & 200   \\
HD~111484~B  & 3-m Lick    & 1998 Jan 21   &  1800 & 280   \\
HD~124257~A  & 3-m Lick    & 1999 May 26   &  2400 & 180   \\
HD~124257~B  & 3-m Lick    & 1999 May 26   &  1800 & 150   \\
HD~124913~A  & 3-m Lick    & 1999 May 27   &  1800 & 110   \\
HD~124913~B  & 3-m Lick    & 1999 May 27   &  1800 & 105   \\
BD+02~2820~A & 3-m Lick    & 1999 May 26   &  2000 &  95   \\
BD+02~2820~B & 3-m Lick    & 1999 May 26   &  2400 & 120   \\
BD+35~2576~A & 3-m Lick    & 1999 May 27   &  2200 & 135   \\
BD+35~2576~B & 3-m Lick    & 1999 May 27   &  2000 & 100   \\
BD+13 2830 A & 3-m Lick    & 2001 March 8  &  2000 &  65   \\
BD+13 2830 B & 3-m Lick    & 2001 March 8  &  2400 &  60   \\
BD+15~2867~B & 3-m Lick    & 1999 May 28   &  2200 &  80   \\
BD+15~2867~C & 3-m Lick    & 1999 May 28   &  2400 &  85   \\
HD 155674    & 3-m Lick    & 1999 May 28   &  2000 &  85   \\
BD+54 1862   & 3-m Lick    & 1999 May 28   &  2000 & 110   \\
BD+65~1043   & 3-m Lick    & 1998 Sept  4  &  1800 & 105   \\
BD+65~1044   & 3-m Lick    & 1998 Sept  4  &  2000 & 100   \\
BD+74~0718   & 10-m Keck I & 1998 Sept 20  &   200 & 95    \\
BD+74~0719   & 10-m Keck I & 1998 Sept 20  &   200 & 150   \\
HD~167215    & 3-m Lick    & 1998 Sept  3  &   800 & 120   \\ 
HD~167216    & 3-m Lick    & 1998 Sept  3  &   800 & 110   \\ 
BD+69~0993~A & 10-m Keck I & 1998 Sept 20  &   120 & 155   \\ 
BD+69~0993~B & 10-m Keck I & 1998 Sept 20  &   120 & 115   \\
BD+08~4386~A & 10-m Keck I & 1998 Sept 20  &   180 & 105   \\ 
BD+08~4386~B & 10-m Keck I & 1998 Sept 20  &   180 & 110   \\ 
BD+03~4428~A & 3-m Lick    & 1998 Sept  4  &  1500 & 120   \\ 
BD+03~4428~B & 3-m Lick    & 1998 Sept  4  &  1800 & 110   \\  
BD+76~0835~A & 10-m Keck I & 1998 Sept 20  &   180 & 160   \\
BD+76~0835~B & 10-m Keck I & 1998 Sept 20  &   180 & 150   \\
BD+01~4575~A & 3-m Lick    & 1998 Sept  4  &  2500 &  80   \\
BD+01~4575~B & 3-m Lick    & 1998 Sept  4  &  2500 &  80   \\
BD+11~5033~A & 3-m Lick    & 1998 Sept  3  &  1200 & 100   \\
BD+11~5033~B & 3-m Lick    & 1998 Sept  3  &  1500 & 105   \\
HD~224984    & 3-m Lick    & 1998 Sept  3  &   800 & 110   \\
HD~224994    & 3-m Lick    & 1998 Sept  3  &  1000 &  90   \\
\enddata
\end{deluxetable}

\begin{deluxetable}{lrrlrrrrrrl}
\tabletypesize{\scriptsize}
\tablecaption{Photometric Data for Program Stars. \label{tbl-1}}
\tablewidth{0pt}
\tablehead{
\colhead{HD/BD} & 
\colhead{$V$} & 
\colhead{$B-V$} &
\colhead{Sp.T.} &
\colhead{$J$} & 
\colhead{$J-H$} & 
\colhead{$J-K$} &
\colhead{$b-y$} & 
\colhead{m1} &
\colhead{c1} &
\colhead{Sources} 
}
\startdata
HD 4552      & 8.87  & 0.53  & F8  &  7.84 & 0.22 & 0.29  &       &       &       &     \\ 
BD+12 0090   & 9.20  & 0.56  & G0  &  8.07 & 0.30 & 0.34  &       &       &       &     \\
HD 6872 A    & 8.06  & 0.41  & F8  &  7.16 & 0.16 & 0.23  & 0.291 & 0.169 & 0.421 & M76 \\ 
HD 6872 B    & 8.53  & 0.87  & F8  &  7.50 & 0.24 & 0.32  & 0.336 & 0.155 & 0.376 & M76 \\
HD 8624      & 8.01  & 0.67  & G0  &       &      &       & 0.424 & 0.234 & 0.330 & O94 \\ 
HD 8610      & 8.01  & 0.60  & F8  &       &      &       & 0.400 & 0.191 & 0.405 & O94 \\
BD+60  269   & 8.56  & 0.48  & F8  &       &      &       &       &       &       &     \\ 
BD+60  271   & 9.08  & 0.49  & F8  &       &      &       &       &       &       &     \\
HD 31208     & 8.20  & 0.83  & K0  &       &      &       & 0.496 & 0.393 & 0.312 & S93 \\
BD+07 754    & 8.38  & 0.89  & K0  &       &      &       & 0.515 & 0.424 & 0.307 & S93 \\
HD 39274     & 8.96  & 0.51  & G0  &  7.89 & 0.31 & 0.34  &       &       &       &     \\ 
HD 39275     & 8.99  & 0.84  & G0  &  7.68 & 0.36 & 0.44  &       &       &       &     \\
HD 54046     & 7.72  & 0.64  & G0  &       &      &       & 0.339 & 0.158 & 0.296 & D84 \\
HD 54100     & 7.72  & 0.53  & F8  &       &      &       & 0.362 & 0.159 & 0.316 & D84 \\
BD+28~1698   & 8.27  & 0.53  & G0  &       &      &       &       &       &       &     \\
BD+28~1697   & 8.26  & 0.50  & G0  &       &      &       &       &       &       &     \\
BD+15~2080   & 9.01  & 0.68  & G5  &       &      &       &       &       &       &     \\
BD+15~2079   & 8.77  & 0.59  & G0  &       &      &       &       &       &       &     \\ 
BD+34 2091 A & 9.71  & 0.51  & G0  &  8.60 & 0.29 & 0.36  &       &       &       &     \\
BD+34 2091 B & 9.79  & 0.49  & G0  &  8.86 & 0.23 & 0.32  &       &       &       &     \\  
HD 92222 A   & 9.40  & 0.70  & G0  &  8.17 & 0.38 & 0.45  & 0.438 & 0.281 & 0.328 & D84 \\
HD 92222 B   & 9.30  & 0.79  & G0  &  8.14 & 0.36 & 0.42  & 0.456 & 0.260 & 0.376 & D84 \\
BD+13 2311 A & 9.32  & 0.46  & F8  &  8.74 & 0.24 & 0.25  &       &       &       &     \\
BD+13 2311 B & 8.8   & 0.6:  & F8  &  8.35 & 0.23 & 0.25  &       &       &       &     \\
HD 98744     & 9.18  & 0.46  & G0  &  8.22 & 0.26 & 0.30  & 0.336 & 0.142 & 0.419 & MM78 \\
HD 98745     & 9.39  & 0.47  & G0  &  8.42 & 0.25 & 0.29  & 0.335 & 0.152 & 0.373 & MM78 \\
HD~111484~A  & 8.77  &       & G0  &       &      &       &       &       &       & CS99 \\
HD~111484~B  & 8.83  &       & G0  &       &      &       &       &       &       & CS99 \\ 
HD 124257~A  & 9.7   & 0.8:  & G5  &  8.10 & 0.22 & 0.29  &       &       &       &     \\
HD~124257~B  & 9.6   & 1.1:  & G5  &  8.68 & 0.25 & 0.33  &       &       &       &     \\
HD~124913~A  & 9.0   & 1.0   & G5  &  8.59 & 0.21 & 0.24  &       &       &       &     \\ 
HD~124913~B  & 9.2   & 1.1   & G5  &  8.92 & 0.17 & 0.20  &       &       &       &     \\
BD+02~2820~A & 10.0  & 0.9   & G5  &       &      &       &       &       &       &     \\
BD+02 2820~B & 10.0  & 0.9   & G5  &       &      &       &       &       &       &     \\ 
BD+35~2576~A & 10.20 & 0.67  & K0  &       &      &       &       &       &       &     \\ 
BD+35~2576~B & 10.77 & 0.61  & K0  &       &      &       &       &       &       &     \\
BD+13 2830 A & 9.8   & 1.1   & K0  &  9.28 & 0.34 & 0.37  &       &       &       &     \\
BD+13 2830 B & 9.9   & 1.1   & K0  &  9.37 & 0.33 & 0.43  &       &       &       &     \\ 
BD+15 2867~B &  9.7  & 1.1   & K0  &  9.15 & 0.39 & 0.46  &       &       &       &     \\
BD+15~2867~C &  9.7  & 1.1   & K0  &  9.16 & 0.40 & 0.49  &       &       &       &     \\
HD 155674    & 8.80  & 1.20  & K0  &       &      &       &       &       &       &     \\
BD+54 1862   & 9.29  & 1.24  & K8  &       &      &       &       &       &       &     \\ 
BD+65 1043   & 8.95  & 0.99  & G5  &  7.19 & 0.46 & 0.60  &       &       &       &     \\
BD+65~1044   & 9.67  & 0.43  & F8  &  8.84 & 0.18 & 0.21  &       &       &       &     \\
BD+74 0718   & 11.1: & 0.87  & G0  &       &      &       &       &       &       &     \\
BD+74~0719   & 10.09 & 0.77  & G5  &       &      &       &       &       &       &     \\
HD~167215    & 8.10  & 0.48  & F8  &       &      &       & 0.357 & 0.148 & 0.373 & D84 \\ 
HD~167216    & 8.23  & 0.48  & F8  &       &      &       & 0.346 & 0.160 & 0.357 & D84 \\ 
BD+69 0993~A & 10.1  & 1.0   & G   &       &      &       &       &       &       &     \\ 
BD+69~0993~B & 10.2  & 1.0   & G   &       &      &       &       &       &       &     \\
BD+08 4386~A & 10.9  & 0.7   & G0  &       &      &       &       &       &       &     \\ 
BD+08~4386~B & 10.9  & 0.4   & G0  &       &      &       &       &       &       &     \\ 
BD+03 4428~A &  9.99 & 0.77  & G5  &       &      &       &       &       &       &     \\ 
BD+03~4428~B & 10.20 & 0.80  & G5  &       &      &       &       &       &       &     \\  
BD+76 0835~A & 9.98  & 0.81  & G   &       &      &       &       &       &       &     \\
BD+76 0835~B & 10.04 & 0.64  & G0  &       &      &       &       &       &       &     \\
BD+01 4575~A & 10.7  & 0.6   & G0  &       &      &       &       &       &       &     \\
BD+01~4575~B & 10.7  & 0.6   & G0  &       &      &       &       &       &       &     \\
BD+11 5033~A & 10.26 & 0.87  & K0  &  8.99 & 0.37 & 0.47  &       &       &       &     \\
BD+11~5033~B & 10.11 & 0.93  & K0  &  8.91 & 0.29 & 0.35  &       &       &       &     \\
HD 224984    &  8.50 & 0.55  & F8  &       &      &       &       &       &       &     \\
HD~224994    &  8.72 & 0.56  & G0  &       &      &       &       &       &       &     \\
\enddata
\tablerefs{
(CS99) Cuypers \& Seggewiss  1999; 
(D84) Duncan 1984; 
(M76) Mechler 1976; 
(MM78) Mechler \& McGinnis 1978; 
(O94) Olsen 1994 ; 
(S93) Schuster, Parrao \& Contreras Mart\'\i nez  1993 }
\end{deluxetable}

\begin{deluxetable}{lrrrrrr}
\tabletypesize{\scriptsize}
\tablecaption{Temperatures and Lithium Abundances of Program Stars. \label{tbl-1}}
\tablewidth{0pt}
\tablehead{
\colhead{HD/BD} & 
\colhead{T$_{\rm eff} (J-K)$} &
\colhead{T$_{\rm eff}$ (fit)} & 
\colhead{[m/H]} &
\colhead{EW (\ion{Li}{1})} & 
\colhead{logN(Li)$_{\rm COG}$} & 
\colhead{logN(Li)$_{\rm fit}$}     
}
\startdata
             & (K)   & (K)  &      & m\AA\    &         &          \\
HD~4552~A    &  6140 & 6000 & -0.2 & $<$2     &  $<$1.7 & $<$1.6   \\ 
BD+12~0090   &  5837 & 5750 & -0.2 & 99$\pm$3 &  2.2    &  2.3     \\
HD~6872~A    &  6543 & 6250 & -0.2 & $<$3     &  $<$2.0 &  $<$1.8  \\ 
HD~6872~B    &  5954 & 6250 & -0.2 & 36$\pm$3 &  2.4    &  2.6     \\
HD~8624      &       & 5250 & -0.2 & 31$\pm$4 &         &  1.6     \\ 
HD~8610      &       & 5500 & -0.2 & 17$\pm$3 &         &  1.6     \\
BD+60~269    &       & 6250 &  0.0 & 49$\pm$4 &         &  2.6     \\ 
BD+60~271    &       & 6250 &  0.0 & $<$3     &         & $<$1.8   \\
HD 31208     &       & 5000 & -0.1 & $<$2     &         & $<$1.0   \\
BD+07 754    &       & 5000 & -0.1 & $<$2     &         & $<$1.0   \\
HD 39274     &  5836 & 6000 &  0.0 & 61$\pm$3 &  2.4    & 2.6      \\ 
HD 39275     &  5305 & 5750 & -0.1 & 79$\pm$3 &  2.1    & 2.5      \\
HD 54046     &       & 6000 & -0.5 & 38$\pm$3 &         & 2.4      \\
HD 54100     &       & 6250 & -0.5 & 39$\pm$3 &         & 2.6      \\   
BD+28~1698   &       & 6000 & -0.2 & 65$\pm$2 &         & 2.7      \\
BD+28~1697   &       & 6000 & -0.2 & 67$\pm$2 &         & 2.7      \\
BD+15~2080   &       & 5750 & -0.2 &  $<$2    &         & $<$0.7   \\
BD+15~2079   &       & 6000 & -0.2 &  9$\pm$2 &         & 1.7      \\
BD+34 2091 A &  5723 & 5750 & -0.1 & 88$\pm$4 &  2.6    & 2.6      \\
BD+34 2091 B &  5954 & 5750 & -0.1 & 63$\pm$3 &  2.6    & 2.5      \\
HD 92222 A   &  5257 & 5500 & -0.1 & 24$\pm$3 &  1.4    & 1.7      \\
HD 92222 B   &  5404 & 5500 &  0.0 &  8$\pm$2 &  1.4    & 1.4      \\
BD+13 2311 A &  6403 & 6500 &  0.0 & 37$\pm$3 &  2.8    & 2.8      \\
BD+13 2311 B &  6403 & 6500 &  0.0 & $<$2     &  $<$1.9 & $<$2.0   \\ 
HD 98744     &  6077 & 6250 & -0.2 & $<$2     &  $<$1.5 & $<$1.5   \\
HD 98745     &  6140 & 6250 & -0.2 & 36$\pm$3 &  2.4    & 2.5      \\
HD~111484~A  &       & 6000 & -0.2 & 64$\pm$3 &         & 2.6      \\
HD~111484~B  &       & 6000 & -0.2 & 92$\pm$3 &         & 2.8      \\ 
HD~124257~A  &  6139 & 6000 &  0.0 & 65$\pm$3 &  2.7    &  2.6     \\
HD~124257~B  &  5954 & 5750 &  0.0 & 11$\pm$3 &  1.9    &  1.7     \\
HD~124913~A  &  6472 & 6500 &  0.0 & $<$4     & $<$2.0  & $<$2.2   \\ 
HD~124913~B  &  6763 & 6500 & -0.1 & $<$4     & $<$2.3  & $<$2.1   \\
BD+02~2820~A &       & 6000 &  0.0 & $<$3     &         & $<$1.5   \\
BD+02~2820~B &       & 5500 &  0.0 & $<$3     &         & $<$1.2   \\
BD+35 2576~A &       & 6000 &  0.0 &  8$\pm$2 &         & 1.8      \\
BD+35~2576~B &       & 6000 &  0.0 &  4$\pm$2 &         & 1.7      \\ 
BD+13 2830 A &  5668 & 5750 & -0.2 & $<$5     & $<$1.3  & $<$1.7   \\
BD+13 2830 B &  5354 & 5750 & -0.2 & $<$3     & $<$1.0  & $<$1.4   \\
BD+15~2867~B &  5210 & 5250 & -0.2 & $<$3     & $<$1.0  & $<$1.0   \\
BD+15~2867~C &  5070 & 5250 & -0.2 & $<$2     & $<$0.8  & $<$0.9   \\
HD 155674    &       & 5250 & -0.2 & $<$5     &         & $<$1.1   \\
BD+54 1862   &       & 5250 & -0.2 & $<$3     &         & $<$1.1   \\ 
BD+65~1043   &  4615 & 5250 &  0.0 & $<$2     &  $<$0.4 &  $<$1.2  \\
BD+65~1044   &  6688 & 6500 &  0.0 & $<$3     &  $<$1.6 &  $<$1.5  \\
BD+74~0718   &       & 5500 & -0.1 & $<$3     &         &  $<$1.2  \\
BD+74~0719   &       & 5500 & -0.1 & $<$3     &         &  $<$1.2  \\
HD 167215    &       & 6500 & -0.2 & 43$\pm$3 &         &  2.8     \\
HD~167216    &       & 6500 & -0.2 & 34$\pm$3 &         &  2.7     \\
BD+69~0993~A &       & 6500 &  0.0 & 30$\pm$3 &         &  2.7     \\
BD+69~0993~B &       & 6500 &  0.0 & 34$\pm$3 &         &  2.7     \\
BD+08~4386~A &       & 6000 &  0.0 & $<$2     &         &  $<$1.6  \\
BD+08~4386~B &       & 5500 &  0.0 & $<$3     &         &  $<$1.2  \\
BD+03~4428~A &       & 5500 & -0.2 & $<$3     &         &  $<$1.2  \\
BD+03~4428~B &       & 5500 & -0.2 & $<$3     &         &  $<$1.4  \\
BD+76~0835~A &       & 6000 & -0.2 & $<$2     &         &  $<$1.5  \\
BD+76~0835~B &       & 5750 & -0.2 & $<$2     &         &  $<$1.3  \\
BD+01~4575~A &       & 5750 & -0.2 & $<$3     &         &  $<$1.2  \\
BD+01~4575~B &       & 5750 & -0.2 & $<$3     &         &  $<$1.6  \\
BD+11~5033~A &  5162 & 5250 & -0.1 & $<$3     & $<$1.0  &  $<$1.2  \\
BD+11~5033~B &  5779 & 5000 & -0.1 & $<$3     & $<$1.4  &  $<$1.1  \\
HD~224984    &       & 6000 & -0.1 & 35$\pm$3 &         &  2.3     \\
HD~224994    &       & 6000 & -0.1 & 33$\pm$3 &         &  2.3     \\
\enddata
\end{deluxetable}

\end{document}